# Highly accurate quantum optimization algorithm for CT image reconstructions based on sinogram patterns

Kyungtaek Jun

Institute of Mathematical Science, Ewha Womans University, Seoul, Republic of Korea
Correspondence: ktfriends@gmail.com

## Abstract

Computed tomography (CT) has been developed as a non-destructive technique for observing minute internal images of samples. It has been difficult to obtain photo-realistic (clean or clear) CT images due to various unwanted artifacts generated during the CT scanning process, along with limitations of back projection algorithms. Recently, an iterative optimization algorithm has been developed that uses the entire sinogram to reduce errors caused by artifacts. In this paper, we introduce a new quantum algorithm for reconstructing CT images. This algorithm can be used with any type of light source as long as the projection is defined. Suppose we have an experimental sinogram produced by a Radon transform. To find the CT image of this sinogram, we express the CT image as a combination of qubits. After the Radon transform of the undetermined CT image, we find the combination of the actual sinogram and the optimized qubits. The global energy optimization value used here can determine the value of qubits through a gate model quantum computer or quantum annealer. In particular, the new algorithm can also be used for cone-beam CT image reconstructions and will be of great help in the field of medical imaging.

## Introduction

Tomography is positioned predominantly as a non-destructive technology. Computed tomography (CT) is a technique that allows non-destructive internal observations of a given sample. CT has been widely used to observe internal structures in biology, archaeology, Geo-science, and materials science [1-6]. In particular, CT is extensively used for medical diagnoses. Electron tomography (ET) is another technique utilized in several fields with different uses. CT is mainly used for samples of a few micrometers or more in size, but ET can be used for far smaller molecular structures (nanometers to angstroms). ET is also widely used to study three-dimensional internal structures in biology and materials science [7-10].

There are three main types of tomographic systems: spiral CT [11], electron tomography (ET) [12], and synchrotron X-ray tomography [13]. Each CT system has been developed to suit the size and characteristics of various samples. The back projection technique used in these tomographic systems can be largely divided into three algorithms; the iteration algorithm [14], fast Fourier transform algorithm [15], or artificial intelligence algorithm [16]. In 2022, Lee and Jun introduced an algorithm with fewer errors than the fast Fourier transform algorithm using the entire sinogram [17]. This algorithm iteratively used an optimization method to reduce the difference between the sinogram generated by the Radon transformation and the sinogram obtained by the projection of the CT image. To reduce the error compared to the existing iterative algorithm, image preprocessing was performed to satisfy the Beer-Lambert law as much as possible [18]. Since this algorithm uses the optimization algorithm for the entire sinogram, it has minimal errors with respect to artifacts that may partially occur [19-24].

In this paper, we introduce a new quantum optimization algorithm. The new quantum optimization algorithm accurately obtains the real internal structure of a sample when two conditions are met: the experimental data is error-free, and when the mathematical projection of the undetermined CT image to make a sinogram matches the x ray projection of a real life sample to an entire sinogram. The algorithm represents the pixels of a CT image in qubits. The algorithm uses projection on the undetermined CT image to create the undetermined sinogram. The original projection data used here uses an experimentally obtained sinogram from a CT system. After that, the algorithm obtains a quadratic unconstrained binary optimization (QUBO) or Ising model through the optimization calculation of the undetermined sinogram and the experimentally obtained sinogram. This model determines the value of all qubits by obtaining a global optimal energy from a gateway quantum computer or quantum annealer. Finally, the determined qubit combination related to the global optimal energy can represent the internal structure of a sample. Our quantum optimization algorithm for CT image reconstruction has three major advantages given certain conditions are met. The first advantage is that our algorithm can be used for CT images of any light source types. One condition for this advantage is that the algorithm can only be used if the projection method used to obtain the experimental data is defined. A second advantage is that it can reconstruct highly accurate CT images, assuming that the projected data satisfies the Beer-Lambert law and there is enough clean data without errors. It also assumes that we have a sufficient number of logical qubits available, and that the projection can be calculated mathematically. In this ideal case, our algorithm can find an exact value for the X-ray mass attenuation of the sample. The final advantage is that the new algorithm is highly resistant to artifacts in the projected images. Since our quantum optimization algorithm calculates the difference for the whole sinogram, it can perform accurate approximations even if an error appears in specific parts. In addition, since motion artifacts can be corrected in CT images, the calculations are easier and more accurate than previous methods of modifying projection images [21-24]. Historically, it was difficult to reconstruct a clean CT image through back projection algorithms that utilized cone-beam CT (CBCT) systems due to the geometric limitations of the light source (X-ray) used. Our new algorithm does not suffer from this issue. Furthermore, our optimization algorithm can approximate a clear image even if the number of projected images is not sufficient [17]. Therefore, we believe that our new quantum algorithm for CT image reconstruction will be of great help in the field of medical imaging. In this paper, we use the Radon transform to test our new algorithm. Quantum optimization calculations are performed in D-Wave Advantage, D-Wave simulator, and IBM Quantum.

## Method

A sinogram is an image created by accumulating projected images of an object according to the projection angle. We introduce a QUBO and Ising models that can represent the entire sinogram. These models can reconstruct CT images with the lowest energy in quantum annealing. Also, it can be implemented through the Quantum Approximate Optimization Algorithm [25] in the gate model quantum computer. A new optimization model is a way to reconstruct a CT image utilizing a quantum computer, assuming that there are enough qubits. The new algorithm should use sinograms for the entire projected image in CBCT system and use sinograms according to each axial level in parallel-beam CT system. In this paper, we introduce a reconstruction method for one axial level to simplify the explanation. Also, we used a Radon transform for projection.

### Energy optimization algorithm for the Radon transform

Consider a space of size $n$ by $n$ that includes the cross section of the Shepp-Logan phantom. Let $\alpha_{ij} = \frac{\mu}{\rho}(i,j)$ is X-ray mass attenuation coefficient [26] and a natural number. We can assume this value as the number at the $(i,j)$ position of the sample (See Fig. 1a). Figure 1b is the sinogram produced by the projections of this sample. The value of each pixel in the sinogram can be expressed as $P(\theta, s)$. Now, we consider a reconstructed image $I$ of size $n$ by $n$ consisting of qubits that can be represented in Fig. 1a (See Fig. 1c). If the maximum of integer-valued pixels is less than $2^{m+1}$, then each pixel in the reconstructed image is represented by one of combinations of qubits and binary numbers in Eq. 1.

$$I_{ij} = \sum_{k=0}^{m} 2^k q_k^{ij} \qquad \text{Eq. 1}$$

Here, $q_k^{ij}$ is 0 or 1, and $I_{ij}$ represent any integer from 0 to $2^{m+1} - 1$.

To apply optimization algorithm onto the experimental sinogram, we use a Radon transform on the undetermined CT image. Let $IP$ be the undetermined sinogram transformed by the CT image $I$. For the projection angle $\theta$, the s-th position of $IP$ is calculated as in Eq. 2.

$$IP(\theta, s) = c_{ij} \sum_{i,j} I'_{ij} \qquad \text{Eq. 2}$$

where $I'_{ij}$ denotes the pixel that affects $IP(\theta, s)$ when the CT image is projected and $c_{ij}$ is the overlapping area when $I'_{ij}$ is projected (See Fig. 2). Using the square of the difference between $P(\theta, s)$ and $IP(\theta, s)$, the optimization model is calculated as following:

$$\left(IP(\theta, s) - P(\theta, s)\right)^2 = \left(c_{ij} \sum_{i,j} I'_{ij} - P(\theta, s)\right)^2 \qquad \text{Eq. 3}$$

$$= \left(c_{ij} \sum_{i,j} \sum_{k=0}^{m} 2^k q_k^{ij} - P(\theta, s)\right)^2 \qquad \text{Eq. 4}$$

$$= c_{ij}^2 \left(\sum_{i,j} \sum_{k=0}^{m} 2^k q_k^{ij}\right)^2 + 2P(\theta, s) c_{ij} \sum_{i,j} \sum_{k=0}^{m} 2^k q_k^{ij} + \left(P(\theta, s)\right)^2 \qquad \text{Eq. 5}$$

In Eq. 5, the second term is linear terms in the QUBO model, and the third term represents a part of the optimization value. The first term without $c_{ij}^2$ is calculated as follows:

$$\left(\sum_{i,j} \sum_{k=0}^{m} 2^k q_k^{ij}\right)^2 = \sum_{i,j,k} 2^{2k} \left(q_k^{ij}\right)^2 + \sum_{i \le i', j \le j', k \le k'} 2^{k+k'+1} q_k^{ij} q_{k'}^{i'j'} \qquad \text{Eq. 6}$$

$$= \sum_{i,j,k} 2^{2k} q_k^{ij} + \sum_{i \le i', j \le j', k \le k'} 2^{k+k'+1} q_k^{ij} q_{k'}^{i'j'} \qquad \text{Eq. 7}$$

To drive Eq. 7 from Eq 6, we can convert the square terms by using $\left(q_k^{ij}\right)^2 = q_k^{ij}$ because of $q_k^{ij}$ is 0 or 1. In the second term of Eq. 7, $i, j,$ and $k$ cannot be equal to $i', j',$ and $k'$ at the same time. We can calculate the first term in Eq. 5 as the sum of linear and quadratic terms as in Eq. 7.

Now we can compare two sinograms $P$ and $IP$. To compute the energy minimization model, we subtract the values for each pixel in the two sinograms and square them.

$$F(\theta, s) = \sum_{\theta=0}^{180-d\theta} \sum_{s=1}^{n} \left((IP - P)(\theta, s)\right)^2 \qquad \text{Eq. 8}$$

where $\theta$ is the projection angle, $s$ is the position of the sensor, and $d\theta$ is the amount of change in the projection angle. Now, $F(\theta, s)$ is expressed in linear terms, quadratic terms excluding constant terms. In the QUBO model, constant terms are excluded. The minimum value of the QUBO model is the opposite sign of the summation of constant terms.

**Energy optimization algorithm for the real X-ray data**

Figure 3 is a flowchart for reconstructing a CT image from X-ray data. To obtain a CT image optimized for the internal structure of a real sample, calibrate the X-ray image. The mathematical projection including the Radon transform is directly proportional to the thickness of the sample's mass attenuation coefficient for each projected location. When the projection does not penetrate the sample, the value of the transmitted position appears as zero. On the other hand, the X-ray projection image has some differences from the projection due to various errors. We need mathematical adjustments to the X-ray image to reduce these errors. First, to represent the empty space, the average value of the empty space is subtracted from the entire X-ray image. The processed X-ray image will show the X-ray density. If the X-ray image has positive or negative deviation, the overall image is modified to satisfy the Beer-Lambert law to be effective in applying the new algorithm. Because the X-ray coherent effect is system-dependent, it would be nice to be able to incorporate this effect into a mathematical projection. If the X-ray image contains high density areas such as metal or pixels that create ring artifacts, it is recommended to remove these areas as well. From now on, the CT image can be reconstructed in a similar way to the one calculated above. First, we create a undetermined CT image in logical qubits. A mathematical projection, such as the X-rays used in the experiment, is applied to this CT image. Compute the quantum optimization model of the sinogram obtained from the experiment and the undetermined sinogram. As introduced in Figure 3, the values of logical qubits are determined through QAOA algorithm or quantum annealing. Finally, we can reconstruct an accurate CT image.

**Result and Implementation**

In this paper, we use the Radon transform for projection to show the results of the new algorithm. We use a 2 by 2 image sample to formulate the QUBO model. Suppose we have an image sample as shown in Fig. 1a. To obtain an exact solution for a 2 by 2 reconstruction image, a sinogram consisting of two projections is required (See Fig.4). We use two qubits for each position. So, the $I_{ij}$ can be expressed as $\sum_{k=0}^{1} 2^k q_k^{ij}$. A sinogram consists of projections with projection angles of 0 and 90 degrees. Now let's formulate the QUBO model.

$$((IP - P)(0,1))^2 = \left(\sum_{k=0}^{1} 2^k q_k^{11} + \sum_{k=0}^{1} 2^k q_k^{21} - \alpha_{11} - \alpha_{21}\right)^2 \quad \text{Eq. 9}$$

$$\begin{aligned}
&= (1 - 2(\alpha_{11} + \alpha_{21}))q_0^{11} + (4 - 4(\alpha_{11} + \alpha_{21}))q_1^{11} + (1 - 2(\alpha_{11} + \alpha_{21}))q_0^{21} \\
&\quad + (4 - 4(\alpha_{11} + \alpha_{21}))q_1^{21} + 2\begin{pmatrix} 2q_0^{11}q_1^{11} + q_0^{11}q_0^{21} + 2q_0^{11}q_1^{21} + \\ 2q_1^{11}q_0^{21} + 4q_1^{11}q_1^{21} + 2q_0^{21}q_1^{21} \end{pmatrix} \\
&\quad + (\alpha_{11} + \alpha_{21})^2
\end{aligned}$$

Eq. 10

When Eq. 10 is calculated from Eq. 9, $(q_k^{ij})^2 = q_k^{ij}$ is used. $((IP - P)(0,2))^2$, $((IP - P)(90,1))^2$, and $((IP - P)(90,2))^2$ can be calculated in a similar way. The QUBO matrix $QM$ is obtained in Eq. 11.

$$QM = \begin{pmatrix}
-4 & 8 & 2 & 4 & 2 & 4 & 0 & 0 \\
0 & -4 & 4 & 8 & 4 & 8 & 0 & 0 \\
0 & 0 & -8 & 8 & 0 & 0 & 2 & 4 \\
0 & 0 & 0 & -12 & 0 & 0 & 4 & 8 \\
0 & 0 & 0 & 0 & -12 & 8 & 2 & 4 \\
0 & 0 & 0 & 0 & 0 & -20 & 4 & 8 \\
0 & 0 & 0 & 0 & 0 & 0 & -16 & 8 \\
0 & 0 & 0 & 0 & 0 & 0 & 0 & -28
\end{pmatrix}$$

Eq. 11

The condition $F(\theta, s) \geq 0$ produces the lowest energy $-(\alpha_{11} + \alpha_{21})^2 - (\alpha_{12} + \alpha_{22})^2 - (\alpha_{21} + \alpha_{22})^2 - (\alpha_{11} + \alpha_{12})^2 = -46$ of the QUBO model. The minimum energy in quantum annealing is -46 and is obtained by the following qubit vector:

$$(q_0^{11}, q_1^{11}, q_0^{12}, q_1^{12}, q_0^{21}, q_1^{21}, q_0^{22}, q_1^{22}) = (0, 0, 1, 0, 0, 1, 1, 1)$$

Eq. 12

In D-Wave Advantage system, the occurrence having the lowest energy is 515 from 1,000 anneals. As a result of quantum annealing, the reconstructed image composed of qubit variables in Fig. 4c has the same values as the sample in Fig. 1a.

QUBO model and Ising model are mathematically equivalent. To convert the QUBO model to the Ising model for the minimum energy model, the following transformation is required.

$$q_i \rightarrow \frac{\sigma_i + 1}{2}$$

Eq. 13

For the QUBO matrix $QM$ in Eq. 11, the Ising matrix $IM$ in Eq. 14 is calculated by applying the transformation in Eq. 13. We obtained Ising matrix using the 'dimod.qubo_to_ising' function provided by D-wave Ocean software.

$$IM = \begin{pmatrix}
3 & 2 & 0.5 & 1 & 0.5 & 1 & 0 & 0 \\
0 & 6 & 1 & 2 & 1 & 2 & 0 & 0 \\
0 & 0 & 1 & 2 & 0 & 0 & 0.5 & 1 \\
0 & 0 & 0 & 2 & 0 & 0 & 1 & 2 \\
0 & 0 & 0 & 0 & -1 & 2 & 0.5 & 1 \\
0 & 0 & 0 & 0 & 0 & -2 & 1 & 2 \\
0 & 0 & 0 & 0 & 0 & 0 & -3 & 2 \\
0 & 0 & 0 & 0 & 0 & 0 & 0 & -6
\end{pmatrix}$$

Eq. 14

D-Wave quantum annealer can calculate the lowest energy for QUBO model or Ising model directly, but in IBM quantum computer using gate model, it is possible to calculate the maximum energy value through QAOA provided by Qiskit. Both the QUBO model and the Ising model are available and have a same absolute optimization value compared to quantum annealer's.

Our second test sample is a 2D Shepp-Logan phantom of $30 \times 30$. We used the binary Shepp-Logan phantom as shown in Fig. 5a to test the QUBO model for CT image reconstruction.

The size and pixel values of the image used here are chosen so that the hybrid solver can find the overall minimum energy for the QUBO model at one time. We apply a Radon transform to this image to get a projection data which is an original sinogram. Since binary numbers can be represented by one qubit, each pixel of an undetermined CT image consists of one qubit. A Radon transform is applied to this CT image to create an undetermined sinogram. We calculate the QUBO model required for CT image reconstruction using the difference between the original sinogram and the undetermined sinogram. The CT image of Fig. 5b was reconstructed using the hybrid solver of the D-Wave system. In order to reconstruct a CT image that is identical to the original image, the minimum value for the QUBO model is $-\sum_{\theta=0}^{180-d\theta} \sum_{s=1}^{n}(P(\theta,s)^2)$. The global minimum energy for this test in quantum annealer is $-225,518.918231688$. We obtained the minimum energy of $-225,518.910675$ using the hybrid solver of the D-Wave system. The CT image reconstructed with the value of qubits determined as the minimum energy is shown in Fig. 5b, exactly the same as the original image. The Shepp-Logan phantom in Fig. 5c is a sample rounded up so that each pixel value is an integer from 0 to 1023. In this case, 10 qubits are required to represent each pixel. Similar to the above, if the minimum energy for this QUBO model is obtained using the hybrid solver, the CT image as shown in Fig. 5d can be reconstructed. The global lowest energy of the QUBO model we are targeting is $-33656657418.458885$, but in the quantum annealer we get $-33653444028.65625$ as the minimum energy. Although the relative error of energy was less than 0.01%, there was a difference between the reconstructed CT image and the original image as shown in Fig. 5c.

Finally, we tested the reconstruction of CT images according to the number of projections with the new algorithm. We used the sinogram made with the Shepp-Logan phantom in Fig. 5a as the original data. The minimum energy expected in a quantum annealer is the negative sign of the sum of the squares of each pixel value in the sinogram. We obtained the lowest energy using a hybrid solver and reconstructed the CT image using qubits. The number of projections used in the sinogram was tested from 30 to 18. From Table 1, we can confirm that the CT reconstruction image is identical to the original image.

## Discussion

Since our new algorithm uses an energy optimization model for the whole sinogram, it has two additional advantages over other algorithms. First, it is not significantly affected by noise and produces good results even if the number of X-ray images is not sufficient. A sinogram is obtained by X-ray transmissions while the sample is rotating. Even if a certain part of the X-ray image has errors, the new algorithm reduces the influence of those errors from other angles of the X-ray image. Second, our new algorithm is not confined by the number of nx projections of an nx×nx sample which typically limits CT image reconstructions. When a CT image has variables corresponding to 〖nx〗^2 for an nx×nx sample, our new algorithm can reconstruct an accurate CT image with fewer than nx number of projections. In order to obtain an accurate CT image using previously proposed algorithms, an nx number of X-ray images of different angles are needed. However, our new algorithm can overcome this limitation. Since this algorithm can reconstruct an accurate CT image using a small number of projections, the radiation received by the sample during CT scanning is reduced. This is one of the important factors for medical CT. We expect this algorithm to play a particularly important role in medical CT image reconstruction. We believe that quantum optimization algorithms will bring great advances in imaging diagnostics using CT images.

Basically, the QUBO model or Ising model can be represented by an upper triangular matrix or a symmetric matrix. The more non-zero numbers in the upper triangular matrix, the more logical qubits are used. In case of D-Wave advantage, about 180 fully connected logical variables are available [27]. This is a number that quantum annealer can calculate when using one qubit variable for each pixel in a CT image of 13 by 13 pixels. Since the size of a commonly used CT image is 500 by 500 or more, at least 250,000 logical qubits are required. The hybrid solver of the D-Wave system provides up to one million variables and 200 million biases. The number of logical qubits used in Fig. 5d is 9,000, showing a relative error of less than 0.01%. Despite the small error when obtaining the minimum energy, the reconstructed CT image still has a difference from the original. However, we believe that if the hybrid solver can obtain a more accurate minimum energy, it will be a huge step forward in medical CT imaging. We believe that a hybrid solver in which a quantum computer and a classical computer work together will change our lives in a shorter time because of the problem of the lack of connectivity between qubits in quantum computers. From our results, we believe that the hybrid solver will not be far from being used for CT image reconstruction. In addition, we will continue to develop mathematical algorithms that allow the QUBO model for CT image reconstruction to better find the global minimum energy.

Our new QUBO model is available in quantum annealers and gated model quantum computers together. But our new algorithm seems more suitable for quantum annealers. The reason is that the circuit depth is too long to obtain the results of the QUBO or Ising model in the gate model quantum computer through QAOA. For example, the IBM quantum computer has low connectivity between qubits and the basic gate does not include an RZZ gate.

## Author Contributions Statement

K. J conceived and designed the research problem. All authors developed theoretical results and wrote the paper.

## Competing financial interests

The authors declare no competing financial and non-financial interests.

## Acknowledgment

This research was supported by the quantum computing technology development program of the National Research Foundation of Korea (NRF) funded by the Korean government (Ministry of Science and ICT (MSIT)) (No. 2020M3H3A111036513 and 2019R1A6A1A11051177). The authors would like to thank Edward Kim and Daniel Kwon for their assistance. The authors would like to thank Hyunju Lee for implementing a basic Python code.

## Figures

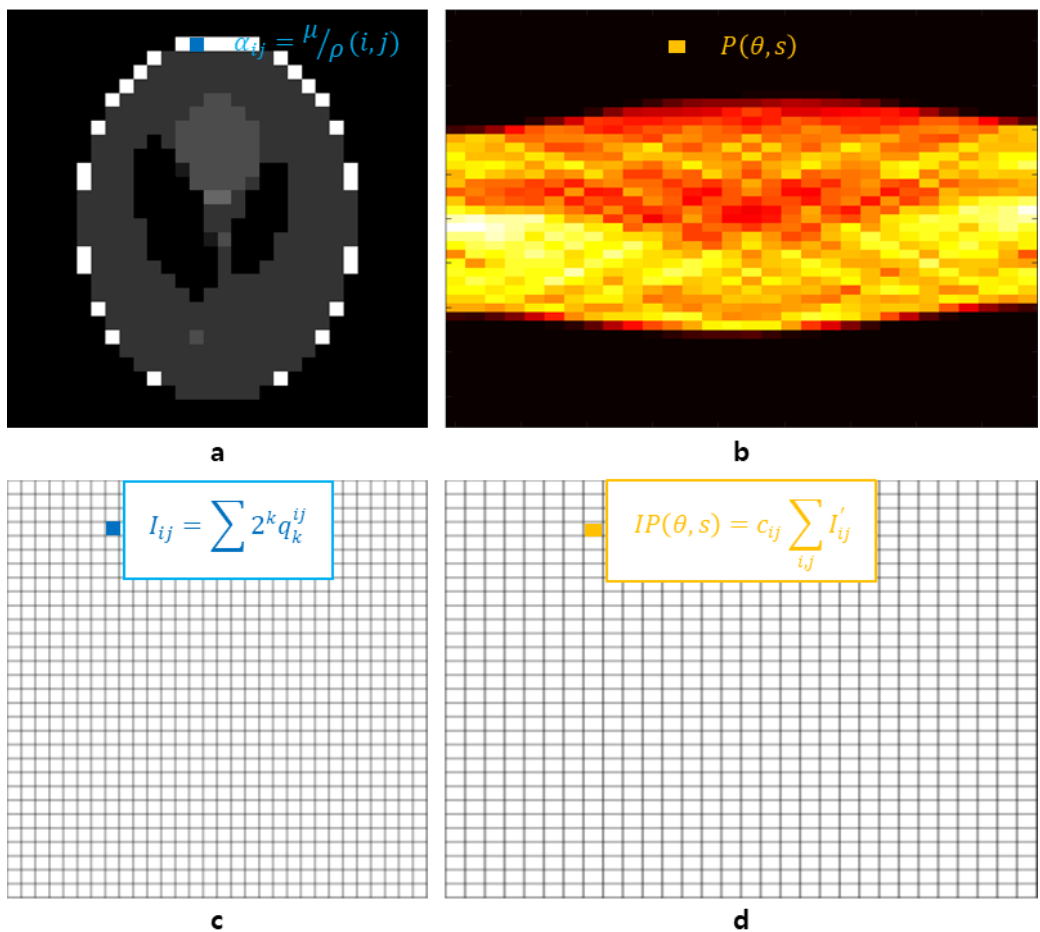

Figure 1. A sample, a CT image, and their two sinograms to illustrate the optimization algorithm. (a) This sample is a $30 \times 30$ Shepp-Logan phantom image. In the case of a general three-dimensional sample, it represents a cross section of the sample corresponding to its axial level. (b) This sinogram was obtained by using Radon transform with the number of pixels equal to the size of the sample in Fig. 1a. When using the data obtained from the CT system, it corresponds to the sinogram of the X-ray image. (c) An undetermined CT image composed of combinations of logical qubits. (d) The sinogram obtained by applying the projection to Fig. 1c in the same way as to obtain the sinogram in Fig. 1b. In this figure, the Radon transform was applied to Figure 1c.

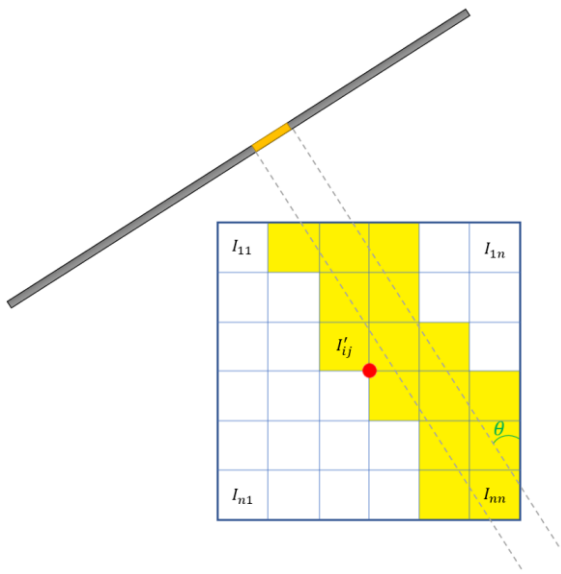

Figure 2. Illustration of the projection of a CT image with respect to the projection angle $\theta$. The $(i,j)$ pixel of the CT image is denoted by $I_{ij}$. Pixels $I_{ij}$ that affect $IP(\theta, s)$ when projected with respect to the projection angle are referred to as $I'_{ij}$.

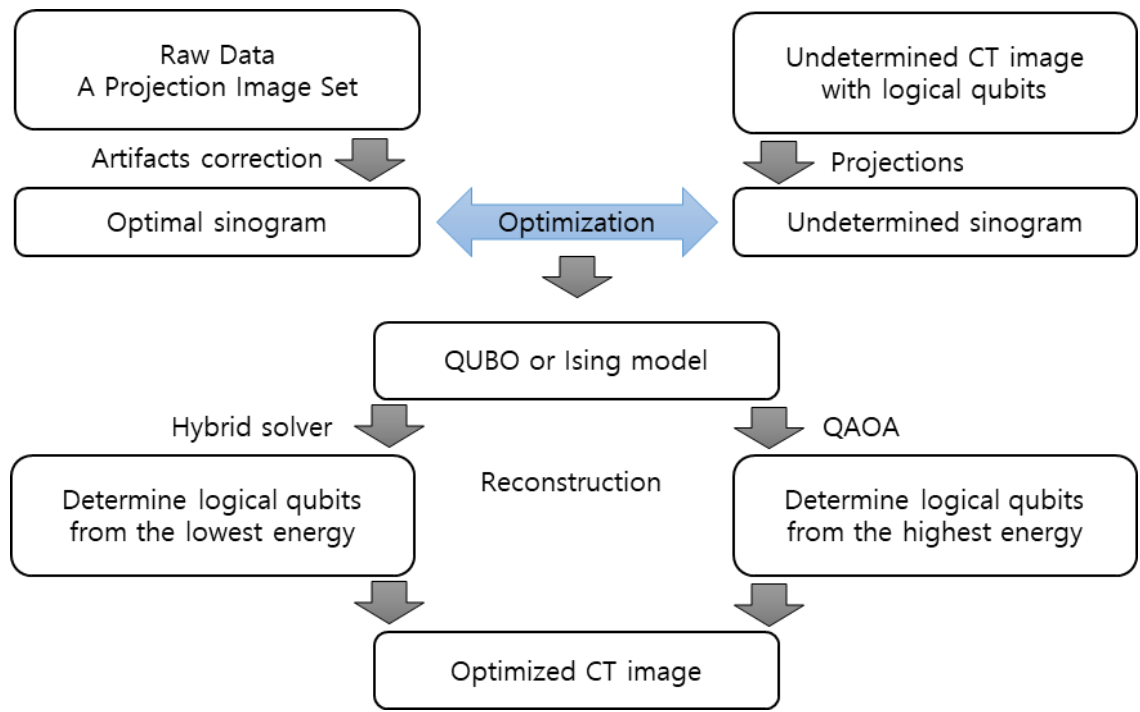

Figure 3. Flowchart of the process of obtaining a CT image. To reconstruct the CBCT image, the entire 3D sinogram is required. To reconstruct the parallel-beam CT image, a sinogram along one axial level are required. In this case, optimization must be applied to all axial levels to obtain 3D CT images.

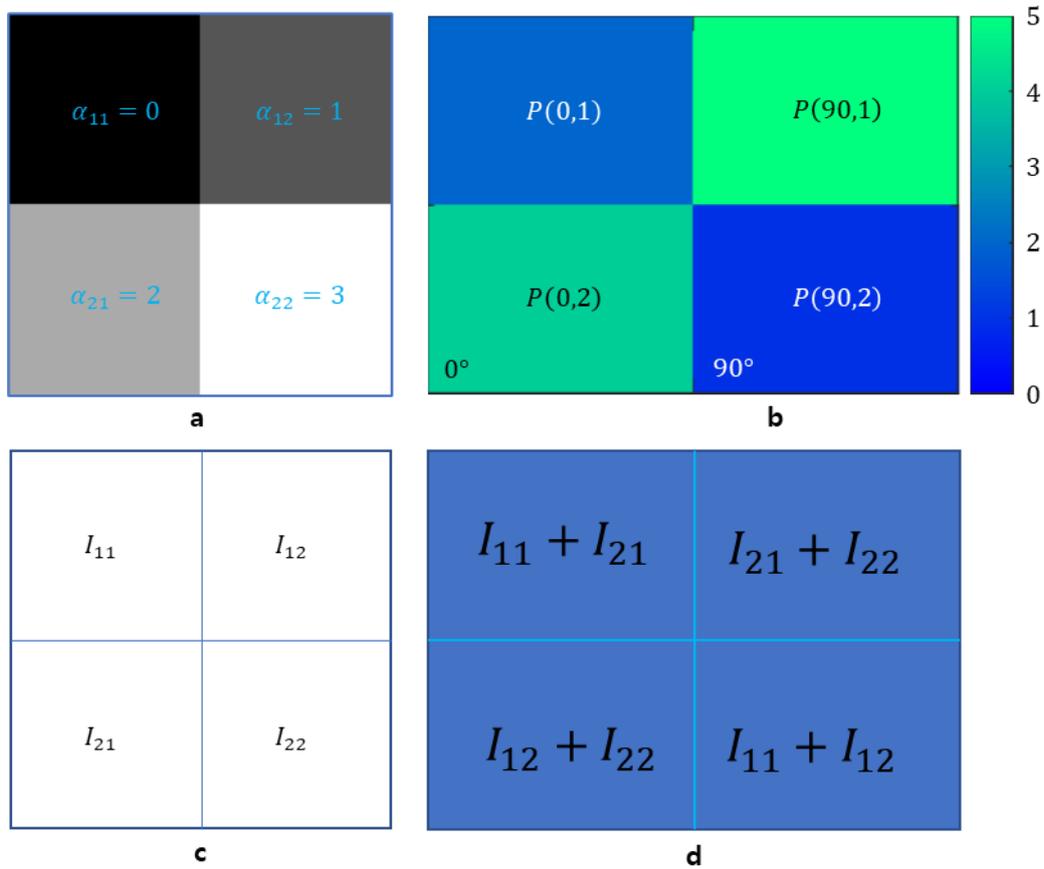

Figure 4. Example used in the QUBO model. (a) 2 by 2 image sample with given mass attenuation coefficients. (b) A sinogram coming from the image sample in Fig. 4a. (c) An undetermined CT image with qubit variables. (d) An undetermined sinogram with two projection angles 0 and 90 degrees by the Radon transform of undetermined CT image in Fig. 4c.

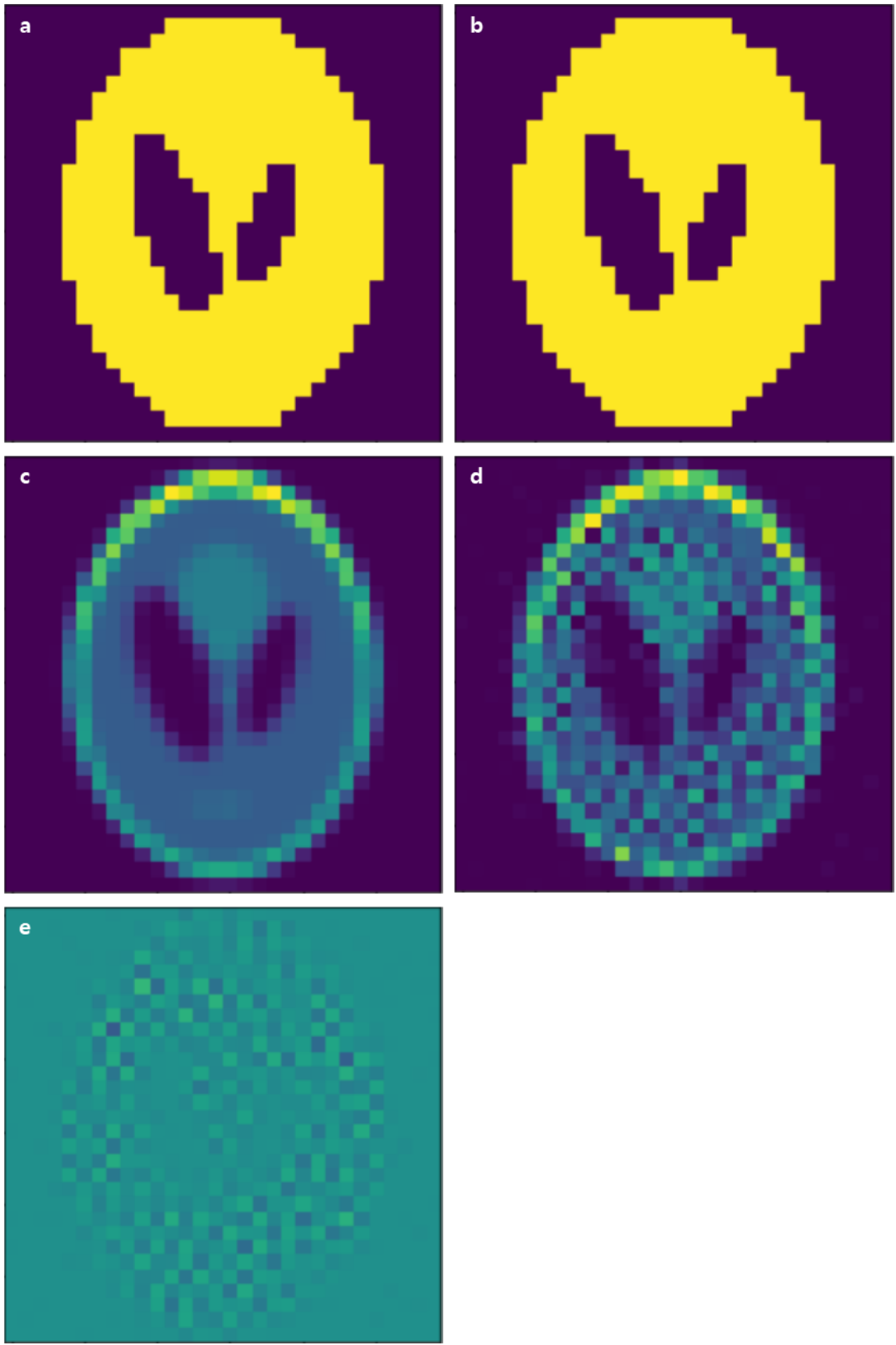

Figure 5. CT images reconstructed by applying quantum optimization algorithm to a Shepp-Logan phantom. The size of the phantom used for testing is 30 × 30. (a) This image is the binary Shepp-

Logan phantom used for testing. (b) This image is a CT image reconstructed using a hybrid solver in the D-Wave system for Fig. 5a. (c) This image is a Shepp-Logan phantom used in the test, and each pixel is rounded to have integer values from 0 to 1023. (d) This CT images are reconstructed using a hybrid solver for Fig.5c. (e) This image shows the difference between the original image in Fig. 5c and the CT image in Fig. 5d. Each pixel darker than the pixels located outside the Shepp-Logan phantom is the case where the CT image has a higher pixel value, and the bright pixels are vice versa.

## Tables

| Projection number | Expected the lowest energy | Result from hybrid solver | Relative Error |
|---|---|---|---|
| 30 | -225518.91823 | -225518.91068 | 0 |
| 27 | -203026.37744 | -203026.37890 | 0 |
| 24 | -180491.95504 | -180491.95515 | 0 |
| 21 | -157920.25285 | -157920.25185 | 0 |
| 18 | -135340.57831 | -135340.58023 | 0 |

Table 1. CT images according to the number of projections used in the new quantum-optimized reconstruction algorithm. In the table, the expected the lowest energy represents the minimum energy that the QUBO model can have. The result from hybrid solver is the lowest energy obtained by using the hybrid solver for the QUBO model. The relative error indicates the difference between the Shepp-Logan phantom and the CT reconstruction image.